\documentclass[12pt]{article}
\usepackage{epsfig}
\usepackage{amsfonts}
\usepackage{color}
\begin {document}

\title {A QUADRUPOLAR GENERALIZATION OF THE EREZ-ROSEN COORDINATES}

\author{J.L. Hern\'andez-Pastora\thanks{E.T.S. Ingenier\'\i a
Industrial de B\'ejar. Phone: +34 923 408080 Ext
2223. Also at +34 923 294400 Ext 1574. e-mail address: jlhp@usal.es}\\
\\
Departamento de Matem\'atica Aplicada
\\Instituto Universitario de F\'\i sica Fundamental y Matem\'aticas. \\ Universidad de Salamanca.  Salamanca,
Espa\~na.  }

\date{\today}

\maketitle

\begin{abstract}

The MSA system of coordinates \cite{msa} for the $MQ$-solution \cite{mq} is proved to be the unique solution of certain partial differential equation with boundary and asymptotic conditions. Such a differential equation is derived from the orthogonality condition between two surfaces which hold a functional relationship.

The obtained  expressions for the MSA system recover the asymptotic expansions  previously calculated \cite{msa} for those coordinates, as well as the Erez-Rosen coordinates in the spherical case. It is also shown that the event horizon of the $MQ$-solution can be easily obtained from those coordinates leading to already known results. But in addition, it allows us to correct a mistaken conclusion related to  some bound imposed to the value of the quadrupole moment \cite{event}.

Finally, it is explored the possibility of extending this method of generalizing the Erez-Rosen coordinates to the general case of solutions with any finite number of Relativistic Multipole Moments (RMM). It is discussed  as well, the possibility of determining the Weyl moments of those solutions from their corresponding MSA coordinates,   aiming to establish a relation between the uniqueness of the MSA coordinates and the solutions itself.

\end{abstract}


\newpage

\section{Introduction}

The Erez-Rosen \cite{R13} system of coordinates (ER) $\{r,y\}$  is specially useful in the context of vacuum solutions of the Einstein equations (\cite{RF2}-\cite{RF5}), and very relevant for describing and interpretation of some axisymmetric solutions of the Weyl family \cite{weyl}:
\begin{eqnarray}
\rho&=&\sqrt{r(r-2M)}\sqrt{1-y^2}   , \qquad r= M + \frac{r_++r_-}{2}\nonumber \\
z&=&(r-M)y   , \qquad \qquad \qquad \quad y= \frac{r_+-r_-}{2M} ,
\end{eqnarray}
where $r_{\pm}\equiv\sqrt{\rho^2+(z\pm M)^2}$, $\{\rho,z\}$ being the standard cylindrical Weyl coordinates.

 First of all, the metric function\footnote{The general solution of the static vacuum Einstein equations for the  axisymmetric case depends on only two metric functions namely  $\Psi$, $\gamma$, and in addition, $\gamma$ is obtained from $\Psi$ by means of a quadrature. Therefore, a solution of these equations is characterized by that metric function $\Psi=\frac 12 \ln (-g_{00})$.  } of the Schwarzschild solution \cite{tesis} becomes extremely simple. While in the ER system of coordinates the metric component  $g_{00}^S=-e^{2\Psi^S}$ simply reads ${\displaystyle g_{00}^S=-1+\frac{2M}{r}}$,  in other systems like the Weyl coordinates $\{\rho,z\}$  the metric function of Schwarzschild solution is rather more complicated:
\begin{equation}
\Psi^S=\frac 12 \ln\left( \frac{r_++r_--2M}{r_++r_-+2M}\right) .
\end{equation}

Secondly, other highlighted comments must be pointed out  about the ER system of coordinates:

1) The general solution of the axisymmetric and static vacuum Einstein equations, for isolated compact objects, is given by the Weyl family in terms of a series
\begin{equation}
\Psi=\sum_{n=0}^{\infty}a_n \frac{P_n(\cos\Theta)}{R^{n+1}} ,
\label{psi}
\end{equation}
where $R\equiv \sqrt{\rho^2+z^2}$ and $z=R\cos\Theta$, $\{R,\Theta\}$ being  the spherical Weyl coordinates.
It is known that the Schwarzschild solution is  described by that series with the following set of non-vanishing Weyl coefficients $a_{2n}^S=-\frac{M^{2n+1}}{2n+1}$. It was proved in \cite{mq} that the sum of the corresponding series (\ref{psi}) for the Schwarzschild solution is  ${\displaystyle \Psi^S=\frac 12 \ln\left(1-\frac{2M}{r}\right)}$, $r$ just being the radial Erez-Rosen coordinate.

2) The expression for the metric component $g_{00}^S$ in that system of coordinates is not only easier than others but it is suitable for being compared with the multipole expansion of the Newtonian gravitational potential. As is known,   the gravitational field of a mass distribution with density $\mu$ involves, in Newtonian Gravity (NG),
solutions of the Poisson equation $\triangle \phi=\mu$, that  can be expanded in a multipolar series
\begin{equation}
\phi=\frac{M}{R}+\frac{M_1^N}{R^2}P_1(\cos\Theta)+\frac{M_2^N}{R^3}P_2(\cos\Theta)+\cdots ,
\label{serieNG}
\end{equation}
where $M_i^N$ denotes the Newtonian Multipole Moments (NMM) ($M_0^N=M$), and $P_i(\cos\Theta)$ are  Legendre polynomials.
Those coefficients $M_i^N$  of the series (\ref{serieNG}) can be written as integrals over the source  \cite{tesis}, and they allow us to characterize the specific solutions constructed by means of  the succession of partial sums of the series (\ref{serieNG}).  In the Erez-Rosen system of coordinates (ER), the quantity ${\displaystyle u\equiv \frac{g_{00}+1}{2}}$ exactly resembles  the form of the gravitational potential $\phi$ in the spherical case.

3) The ER system of coordinates is adapted to the spherical symmetry that  the Schwarzschild solution is representing, since its metric function $\Psi^S$, written in these coordinates, does not depend on the angular variable.

4) Finally, the radial coordinate can be defined univocally as the unique solution of a differential equation  with boundary conditions \cite{msa} which arises from  a symmetry of the Laplace equation  and the  Ernst equation \cite{ernst} as well. In  \cite{Nsym}, the existence of some kinds of  symmetries in Newtonian Gravity (NG) has been proved, which  makes it possible to extract from all solutions of the axially symmetric Laplace equation those with the prescribed Newtonian Multipole Moments (NMM).
A family of vector fields that  are the infinitesimal generators of certain one-parameter groups of transformations can be constructed. These vector fields represent symmetries of certain  systems of differential equations whose
group-invariant solutions   turn out  to be the family of axisymmetric gravitational potentials with  specific gravitational multipoles.

Several works \cite{varios}, \cite{mq}, \cite{sueco} have been devoted  to find  solutions of the vacuum axisymmetric Einstein equations with a finite number of RMM   \cite{geroch}. In particular, the metric function of the $MQ$-solution  \cite{mq} can be described as a series in a dimensionless quadrupolar parameter $q\equiv Q/M^3$, $\Psi=\Psi_0+\Psi_1 q +\Psi_2 q^2+\cdots$ where $\Psi_0=\Psi^S$ denotes the Schwarzschild solution and $Q$ is the quadrupole relativistic moment.
Each one of the functions $\Psi_i$ are known but the sum of the convergent series is not obtained yet. The quantity $u$ corresponding to metrics with a finite number of RMM is wanted to resemble, in some system of coordinates, the gravitational potential $\phi$ (\ref{serieNG}) except for changing NMM by RMM. The possibility of  extrapolating  the symmetries obtained in NG \cite{Nsym} to GR, as well as  characterizing the solutions with a finite number of RMM by means of group-invariant solutions, are the relevant features of a proposed system of coordinates.
In \cite{msa} an answer to these questions was sought by introducing a family of coordinate systems referred to as  MSA ({\it Multipole-Symmetry Adapted}).  The MSA system of coordinates \cite{msa} were introduced to generalize the ER coordinates with the aim of inheriting its benefits in describing the solution with a finite number of RMM analogously to the Schwarzschild case.

The MSA system of coordinates is one of ACMC type. The  system of coordinates called ACMC ({\it Asymptotically Cartesian and Mass Centered}) were introduced in 1980 by Thorne in the context of Multipole expansions of gravitational radiation \cite{thorne}. His work presents a definition of Relativistic Multipole Moments (RMM) and shows us how to deduce the RMM of a source from the form of its stationary and asymptotically flat vacuum metric in an ACMC coordinate system; if the components of the metric are written in these coordinates $\{\hat t, \hat r,\hat\theta, \hat\varphi\}$ we can read off the RMM from the resulting expressions,  other terms,  $R_{ij}^{(n-1)}(\hat y)$,  called {\it Thorne rests}, appearing at the same time. For the case of axial symmetry, the $g_{00}$ component of any static metric written in that kind of coordinates is:
\begin{equation}
g_{00} =-1+\frac2{c^2} \left[ \sum_{n=0}^\infty \frac1{\hat r^{n+1}}
M_{n} P_{n}(\hat y)+ \sum_{n=1}^\infty \frac1{\hat r^{n+1}}
R_{00}^{(n-1)}(\hat  y) \right] , \label{g00thorne}
\end{equation}
$M_{n}$ being the RMM of order $n$, and $P_n(\hat y)$ the Legendre polynomial with $\hat y\equiv \cos\hat\theta$.

There exists a  gauge freedom in the choice of ACMC coordinates preserving the invariance of the first series in (\ref{g00thorne}) and addressing the differences in the metric expansion to the Thorne rests. Among the broad class of coordinate systems of this type, we look for the system  leading to an expansion of the metric in such a way that all the $R_{00}^{(n-1)}(\hat y)$ Thorne rests vanish. This fact is the principal feature of the MSA system and its specific definition which make it  to be unique since  only one asymptotically cartesian system of coordinates leads to vanishing Thorne rests.

By introducing these coordinates, the function $u$ associated to an  static and axially symmetric vacuum solution with a finite number of RMM, should satisfy the same system of differential equations as the classical potential in NG, and  the symmetries of these equations \cite{Nsym} thus allow us to describe and determine the Multipole Solutions in GR analogously to the Newtonian case.

In \cite{msa}, the MSA coordinates for the $MQ$-solution were obtained by means of an asymptotic expansion, and in addition the existence and uniqueness of such coordinates were tried to be established from the resolution of a Dirichlet problem. While it is true that the aim for the spherical case was successful, nevertheless outcome cannot be obtained for the general case because no suitable boundary condition was available. Now, the aim of this work is to show that MSA coordinates for the $MQ$-solution can be explicitly  calculated   and not only in the form of an asymptotic expansion at infinity. In addition, we shall show that these coordinates will be the unique solution of  certain differential equation preserving a set of asymptotic and boundary conditions. As we shall explain, the differential equation arises from the orthogonality condition of the coordinates and the functional relation among them.

This work is organized as follows.
In Section 2  the MSA system of coordinates is defined  and the problem to solve is posed. Several strategies for their resolution are explained and the spherical case is broadly treated. Nextly the resolution of MSA coordinates for the $MQ$- solution is considered. The existence of a functional relation between the radial and angular variables  leads to establish a different statement of the problem to solve. Finally, this section is devoted to solve from this point of view the new equations set up for the MSA coordinates.

Section 3 boards the relevance and more important aspects of the MSA coordinates found. In particular, the asymptotic expansion recovers the results previously obtained in \cite{msa}, and the spherical limit leading to the ER coordinates is shown.  An application to the calculation of the event horizon of the solution is discussed and mistaken conclusion from previous work is corrected. It also
contains  some comments about  possible future generalizations.

A conclusion section summarizes  the aims achieved.

Finally, two appendices are included where related calculations from the previous sections  are addressed: appendix 1 contains a proof for a theorem  allowing to solve certain type of differential equation involving both the Laplacian and the  gradient operators, and appendix 2 is devoted to complete the calculation of the functions
involved in the construction of the solution for the {\it statement 2} in the general case.

\section{The MSA system of coordinates}

\subsection{Definition and problem statement}

We are looking for a change of coordinates  that allows us to write the $g_{00}$ metric component of the solutions with a finite number of RMM as follows
\begin{equation}
g_{00}^{RMM}=-1+2 \Phi , \qquad \Phi\equiv \sum_{i=0}^N \frac{M_{2i}}{\hat r^{2i+1}}
P_{2i}(\hat y)
\label{g00}
\end{equation}
where $M_{2i}$ represent the massive RMM of the given solution with a finite number $N$ of multipoles, and $P_{2i}(\hat y)$ denote the Legendre polynomials, $\{\hat r,\hat y\}$ being such a system of coordinates that has been called MSA ({\it Multipole Symmetry Adapted}).

The  metric component $g_{00}^{RMM}$ must satisfy the Ernst equation in the new system of coordinates which, in addition, is required to preserve orthogonality. Hence, three conditions are imposed on the system of coordinates, let us recall them:
(i)the desired form (\ref{g00}) of the  metric component $g_{00}^{RMM}$, (ii) the fulfilment of the Ernst equation, and (iii)  the orthogonality condition. This set of conditions is redundant, and this fact was already discussed in \cite{msa} leading to conclude that the pair of conditions (i)-(ii) is equivalent to the pair (i)-(iii). We shall back to show this redundancy later.

Therefore, the problem that we want to solve could be formulated as follows:

{\it Statement 1}

Let us find a couple of functions $\hat r=\hat r(x_i)$, $\hat y=\hat y(x_i)$,  $\{x_i\}$ being the initial coordinates in\footnote{The axial and stationary symmetries of the solutions we want to handle imply the existence of two cyclic coordinates (time coordinate and axial angle). Therefore the change of coordinates is restricted to a plane containing the symmetry axis.} $\mathbb R^2$ verifying the following equations:
\begin{eqnarray}
& (i) & g_{00}^{RMM}(\hat r,\hat y)=-1+2\Phi  \nonumber \\
& (ii) & (\nabla \hat r )^2 a_1(\hat r, \hat y) + (\triangle \hat r) a_3(\hat r, \hat y) =(\nabla \hat y )^2 a_2(\hat r, \hat y) + (\triangle \hat y) a_4(\hat r, \hat y)  \nonumber \\
& (iii) & (\nabla \hat r) (\nabla \hat y) =0\ ,
\label{p1}
\end{eqnarray}
where $\nabla$, $\triangle$ denote the gradient and Laplacian operators respectively with respect to an Euclidean metric (see \cite{msa} for details), and
\begin{eqnarray}
a_1(\hat r, \hat y)& \equiv &\partial_{\hat r \hat r} \Phi-\frac{2}{2\Phi -1}(\partial_{\hat r} \Phi)^2, \qquad a_3(\hat r, \hat y) \equiv \partial_{\hat r} \Phi  \nonumber \\
a_2(\hat r, \hat y)& \equiv &\partial_{\hat y \hat y} \Phi-\frac{2}{2\Phi -1}(\partial_{\hat y} \Phi)^2, \qquad a_4(\hat r, \hat y) \equiv \partial_{\hat y} \Phi
\label{aes}
\end{eqnarray}
In addition we demand  of the new coordinates a Cartesian asymptotic behaviour, i.e. $\hat r$, $\hat y$ are required to recover,  at infinity, the spherical radial coordinate $R$ and the angular variable ($\omega\equiv \cos\Theta$) associated to Cartesian coordinates respectively: $\hat r (R\rightarrow\infty)=R+\alpha$, $\hat y(R\rightarrow\infty)=\omega$, $\alpha$ being an arbitrary constant. This condition is needed to guarantee that the new coordinates are ACMC type.

\subsection{Resolution strategies}

In \cite{msa}, a coordinate transformation with a good asymptotic Cartesian behaviour  was performed as follows
\begin{eqnarray}
\hat r=R\left[1+\sum_{n=1}^{\infty}f_n(\omega)\frac{1}{R^n}\right] \nonumber \\
\hat y=\omega+\sum_{n=1}^{\infty}g_n(\omega)\frac{1}{R^n} \ ,
\label{transf}
\end{eqnarray}
and so, asymptotic expansions of the new coordinates were considered from a starting point. Next, the unknown functions $f_n(\omega)$ and $g_n(\omega)$ were obtained from the conditions (i)-(iii). It was also discussed that there exist two different ways to implement those conditions: on the one hand we can introduce expressions (\ref{transf}) into the orthogonality condition (iii) leading to relations between both sets of functions $f_n$, $g_n$ (see equation (56)  in \cite{msa} and comments therein for details\footnote{Let us note  a misprint in that equation, since the equal symbol and the minus one before the second sum should exchange their positions.}.) Next, the condition (i)univocally determine all the functions $f_n$, $g_n$ at any order of the expansion series.
On the other hand, identical results are obtained  if the gauge (\ref{transf}) is implemented into the pair of conditions (ii) and (iii).

Arguments about the convergence of the series (\ref{transf}) are also shown in \cite{msa}. Nevertheless, this gauge transformation  is purely an asymptotic approach to the new system of coordinates since we are not able to obtain the sum of those series, and any attempt to cut the series at some order cannot guarantee accuracy of the gauge except for the case of being at  large distances from the source.

In this work we want to tackle the issue  to figure out the problem from the  {\it statement 1}, without considering the asymptotic approach. The particular case of the MSA coordinates associated to the vacuum solution of the axisymmetric static Einstein equations with the mass $M$ as its unique RMM deserves a detailed discussion. As is well known those coordinates are the so-called {\it standard coordinates of Schwarzschild} which allow us to write the metric component of the Schwarzschild solution as ${\displaystyle g_{00}^{S}=-1+\frac{2M}{\hat r}}$. Of course that this case is specially simple, because just only the condition (i) by itself univocally determines   the radial coordinate $\hat r$. Since the Schwarzschild solution  represents the spherical symmetry and $\Phi=2M/\hat r$ does not depend on the angular variable $\hat y$ we can solve from condition (i) the radial coordinate $\hat r$ as follows:
\begin{equation}
\hat r =\frac{2M}{g_{00}^S+1} \ ,
\label{rdeschw}
\end{equation}
and the well known expression $\hat r=M(x+1)$ is obtained if we use prolate spheroidal coordinates $\{x,y\}$ \cite{prolates} as initial coordinates, since the metric component of the Schwarzschild solution is  ${\displaystyle g_{00}^S=-\frac{x-1}{x+1}}$. The angular variable $\hat y$ must be  orthogonal to $\hat r$ (condition (iii)), i.e. the prolate angular variable $y$.

Getting back to the asymptotic procedure described in \cite{msa}, we want to remind that,  in this particular case, it is possible to obtain the sum of the series describing the gauge of coordinates (see \cite{msa} for details) and the standard coordinate of Schwarzschild is recovered. But, in addition, it is not necessary to perform the asymptotic expansion (\ref{transf}) of the new coordinates to solve the {\it statement 1}  since condition (ii) can be integrated in this particular case. In fact, existence and uniqueness of the problem  with only one RMM can be deduced with the help of the boundary condition $\hat r(y=\omega=\pm 1)=R+M$ on the symmetry axis. Let us briefly see the procedure; for this spherical case we have $\Phi=M/\hat r$ and so, $a_2=a_4=0$ from (\ref{aes}) which imply that condition (ii) leads to the following equation
\begin{equation}
\frac{2(M-\hat r)}{\hat r (2M-\hat r)}(\nabla \hat r)^2=\triangle \hat r ,
\end{equation}
or equivalently
\begin{equation}
H \triangle  \bar r-\frac{d H}{d \bar r} (\nabla \bar r)^2=0 ,
\end{equation}
where $H \equiv\bar r^2-M^2$ with $\bar r\equiv \hat r-M$. By virtue of Theorem 1 appearing in the appendix, this equation  admits the solution $\bar r=\bar r(\xi)$,    $\ \xi$ being an harmonic function  ($\triangle \xi=0$), {\it iff} ${\displaystyle H=k\frac{d \bar r}{d \xi}}$ for any constant $k$, and  hence we have that
\begin{equation}
\bar r=M\left(\frac{1+e^{\frac{2M \xi}{k}}}{1-e^{\frac{2M \xi}{k}}}\right) , \quad  \hat r=\frac{2M}{1-e^{\frac{2M \xi}{k}}} .
 \label{rdexi}
\end{equation}
The auxiliary harmonic function $\xi$ (as well as the radial coordinate $\hat r$) depends on the initial coordinates we were using; if we start with standard Weyl coordinates $\{R,\omega\}$ the boundary condition $\hat r(y=\omega=\pm 1)=R+M$ on the symmetry axis determines the auxiliary function $\xi$, since $\xi$ being harmonic
\begin{equation}
\xi=\sum_{n=0}^{\infty}\frac{b_n}{R^{n+1}}P_n(\omega) \ ,
\end{equation}
and the coefficients $b_n$  must be equal to the so called Weyl moments $b_{2n}=a^S_{2n}=-\frac{M^{2n+1}}{2n+1}$, $b_{2n+1}=a^S_{2n+1}=0$ in order to fulfill that boundary condition. As is known, these Weyl moments correspond to those coefficients of the Schwarzschild metric function $\Psi^S$, where $g_{00}^S=-e^{2\Psi^S}$. Alternatively, an harmonic  function $\xi$, in prolate spheroidal coordinates, such that the expression (\ref{rdexi}) leads (with $k=M$) to the known  radial Schwarzschild coordinate $\hat r=M(x+1)$ is no other than $\xi^S=\frac 12 \ln\left(\frac{x-1}{x+1}\right)$. Hence, whatever the coordinates we were using,  we can see from the equation (\ref{rdexi}) that the auxiliary harmonic function $\xi$ must be in fact the metric function $\Psi^S$ of the Schwarzschild solution, or equivalently,
$e^{\frac{2M \xi^S}{k}}=-g_{00}^{S}$ and  the equation (\ref{rdeschw}) for the radial coordinate $\hat r$ is just (\ref{rdexi}).

For the general case, the condition (ii) is quite difficult to solve since it is a non-linear partial differential equation for two variables. Nevertheless, the combination of conditions (i) and (ii) leads to a simplification of the problem. This is the new strategy to solve (ii) that we shall consider from now onwards. In what follows, at least for the moment, we limit ourselves to the case of the $MQ$-solution.

Let us make use of the condition (i) to obtain a functional relation between $\hat r$ and $\hat y$, as follows:
\begin{equation}
\hat y  = \frac{1}{3Q}\sqrt{F \hat r^3-2 M \hat r^2+Q}
\label{yder}
\end{equation}
where $F\equiv1+g_{00}(x^i)$ denotes that function written in the system of coordinates $\{x^i\}$. We use  those coordinates with which  the explicit form of the  $MQ$-metric is known. In \cite{mq} we obtained that solution in prolate spheroidal coordinates. Now, by taking the expression (\ref{yder}) either into condition (ii) or into the orthogonality condition (iii) the following equation is obtained:
\begin{equation}
(-3F \hat r+4 M)(\nabla \hat r)^2=\hat r^2 \nabla \hat r \nabla F
\label{3rF}
\end{equation}
and hence, a solution to the above equation (\ref{3rF}) fulfilling the asymptotic condition and with the appropriate boundary behaviour at the symmetry axis $\hat r(x,y=\pm1)\equiv \hat{r_1}$
would be a suitable answer to the problem of the  {\it statement 1}. The boundary condition at the axis of symmetry is given from the condition (i):
\begin{equation}
F(x,y=\pm1)\equiv F_1=\frac{2M}{\hat r_1}+\frac{2Q}{\hat r_1^3}
\label{F1}
\end{equation}
We shall back to the explicit expression for $\hat r_1$ later, but previously let us redefine the problem to solve enunciated at {\it Statement 1}.
In accordance with the discussion in \cite{msa} we have showed that conditions (i)-(ii) lead to the same equations that the pair of conditions (i)-(iii), and therefore the problem we want to solve is now formulated as follows:

{\it Statement 2}

Let  $\hat r=\hat r(x,y)$, $\hat y=\hat y(x,y)$ be  a pair of surfaces with the relationship given by the expression (\ref{yder}). Then, the radial coordinate of the MSA system of coordinates $\{\hat r, \hat y\}$ with which the $MQ$-solution is described by the metric function ${\displaystyle g_{00}^{MQ}=-1+\frac{2M}{\hat r}+\frac{2Q P_2(\hat y)}{\hat r^3}}$, is given by the solution of the differential equation (\ref{3rF}) with the boundary condition $\hat r(x,y=\pm1)=\hat r_1$ and asymptotic conditions  $\hat r (R\rightarrow\infty)=R+\alpha$, $\hat y(R\rightarrow\infty)=\omega$.

\subsection{A solution for the statement 2}

Let us try a change of the radial coordinate  $\hat r$, supposing, for now, that it depends only in a new function $\zeta$: $\hat r=\hat r(\zeta)$. Then, equation (\ref{3rF}) becomes:
\begin{equation}
(-3F \hat r+4 M)(\nabla \zeta)^2 (\hat r^{\prime})^2=\hat r^2 \hat r^{ \prime} \nabla \zeta \nabla F ,
\label{3rFnew}
\end{equation}
where $\hat r^{\prime}$ denotes  the derivative with respect to $\zeta$.

Now we make the following ansatz:
\begin{equation}
(-3F \hat r+4 M) \hat r^{\prime}=\hat r^2 ,
\label{ansatz}
\end{equation}
and consequently from equation (\ref{3rFnew}) we have that
\begin{equation}
(\nabla \zeta)^2 = \nabla \zeta \nabla F .
\label{planos}
\end{equation}
We shall discuss the reason for using this ansatz, but previously let us see the solutions of the pair of equations (\ref{ansatz})-(\ref{planos}). The trivial solution of  equation (\ref{planos}) is $\zeta =F$, but $\zeta=F-G$ is also a solution  whatever function $G$ being  again a solution  itself of equation (\ref{planos}). Nevertheless, in what follows, we shall deal with  the case $G=0$, addressing the discussion  on the general case to the appendix 2. With respect to the ansatz (\ref{ansatz}) it is a linear equation for $\zeta=\zeta(\hat r)$ whose general solution (with $F=\zeta$) is the following
\begin{equation}
\zeta= \frac{2M}{\hat r}+\frac{C(x,y)}{\hat r^3}=F .
\label{zetader}
\end{equation}
This expression (\ref{zetader}) is the main reason for using the ansatz (\ref{ansatz}):\  the radial coordinate on the axis is obtained when the arbitrary function becomes constant $C(x,y=\pm1)=2Q$ (see equation  (\ref{F1})), and so the boundary  condition on the axis from {\it statement 2} is included in the solution.   This boundary condition  is needed to fulfil at the same time  the equation (\ref{zetader}) as well as the condition (i) (\ref{p1}). In fact,  the function $C(x,y)$ cannot be  a constant except on the symmetry axis and hence the radial coordinate $ \hat r$ will no longer be an exclusive function of $\zeta$ but  also dependent on the initial coordinates throughout the function $C(x,y)$, i.e. $\hat r=\hat r(\zeta, C\{x,y\})$. Therefore, at this point we need to recalculate  equation (\ref{3rFnew}) from equation (\ref{3rF}) to determine the function $C(x,y)$, but first let us see what can we say about that function; by taking into account the condition (i) and the equation (\ref{zetader}) we have that
\begin{equation}
C(x,y)=P_2(\hat y) 2 Q ,
\label{CP2Q}
\end{equation}
or equivalently, the angular coordinate $\hat y$ from (\ref{yder}) has to be
\begin{equation}
\hat y=\sqrt{\frac{C(x,y)+Q}{3 Q}} .
\label{ydeC}
\end{equation}
Let us note that the condition imposed to $C(x,y)$ on the symmetry axis $C(x,y=\pm 1)=2Q$ is also the guarantee  to preserve the symmetry axis in the change of coordinates, i.e., $\hat y(x,y=\pm1)=y=\pm 1$. Hence, with all  these considerations, we can state for that function the following behaviour
\begin{equation}
C(x,y)=2Q+\Pi(x,y), \ \Pi(x,y=\pm 1)=0
\label{CPi}
\end{equation}
In addition, since we want to obtain in the spherical case ($Q=0$) the Schwarzschild coordinates ($\hat y=y$ for the angular coordinate), then the function $\Pi(x,y)$ must be  at least of order $Q$ as follows
\begin{equation}
\Pi(x,y)=-3Q(1-y^2)+\sum_{i=2}^{\infty}
q^i c_i(x,y),
\label{Pi}
\end{equation}
with $q\equiv \frac{Q}{M^3}$ the dimensionless quadrupolar parameter and $c_i(x,y=\pm 1)=0$. From (\ref{ydeC}) we have finally
\begin{equation}
\hat y=\sqrt{ y^2+\frac{1}{3M^3}\sum_{i=2}^{\infty}
	q^{i-1} c_i(x,y)} .
\label{ydeces}
\end{equation}

Let us back to clarify two subjects before mentioned and still unsettled. In the one hand we have to obtain  the explicit expression of the radial coordinate $\hat r$  from (\ref{zetader}), and in the other hand we have to obtain a differential equation for the function $C(x,y)$ since equation (\ref{3rFnew}) is no longer compatible with equation (\ref{3rF}) and $\hat r=\hat r(\zeta, \{x,y\})$. With respect to  $\hat r$ we have to find the existing real root of equation  (\ref{zetader}), which can be expressed  as the following piecewise function
\begin{equation}
	\hat r=\frac{a Z}{3 F}, \ Z \equiv \left\{
	\begin{array}{cll}
H+\frac 1H +1, & H\equiv \left(\chi+\sqrt{\chi^2-1}\right)^{1/3} &,  \mid \chi \mid \geq 1 \nonumber\\
1+\cos(2\varphi), & \varphi\equiv \frac 13 \arccos(\chi) & , \mid \chi \mid \leq 1,
	\end{array}
	\right.
\label{rdeZ}
\end{equation}
where ${\displaystyle \chi\equiv\frac{27}{2}\frac{CF^2}{a^3}+1}$, and $a\equiv 2M$. $Z$ is continuous at $\chi=1$ where ($Z(\chi=1)=3$) $\hat r(\chi=1)=a/F$ and it corresponds to either a surface independent of the angular variable in the spherical case when $q=0$ ($C=0$) or the radial coordinate $\hat r$ at a fixed value of the angular coordinate $\hat y=\pm \frac{1}{\sqrt 3}$ in the general case. The other value $\chi=-1$ is not available for the piecewise function (\ref{rdeZ}), in fact $\chi >-1$ since $F \in[1,0)$, the Legendre polynomial $P_2 \in [-1/2,1]$ , and $\mid q\mid <1$ (we make use of the relation (\ref{CP2Q})). And finally, the boundary value for $\hat r$ on the symmetry axis, $\hat r(x,y=\pm 1)\equiv \hat r_1$ is
\begin{equation}
\hat r_1=\frac{a Z_1}{3 F_1}
\end{equation}
where $Z_1$ denotes the function $Z$ from (\ref{rdeZ}) with ${\displaystyle \chi(x,y=\pm1)=\frac{27}{8} q F_1^2+1}$.

On the other hand we now come to the determination of the differential equation for the function $C$. Let us remind that the proposal $\hat r=\hat r(\zeta)$ introduced into  equation (\ref{3rF}) leads to equation (\ref{3rFnew}). Since we have shown that the assumed ansatz (\ref{ansatz}) requires $\hat r=\hat r(\zeta,\{x,y\}))$ then  equation (\ref{3rF}) becomes
\begin{equation}
\hat \nabla \hat r \cdot \nabla F =-\frac{1}{\hat r^{\prime}}(\hat \nabla \hat r)^2
\label{newplanos}
\end{equation}
instead of (\ref{3rFnew}),  where the ansatz (\ref{ansatz}) has been used again.  The symbol ($\ \hat{}\ $) over the gradient operator $\hat \nabla$ denotes that it strictly acts   on the function $C(x,y)$.
By making use of (\ref{zetader}) to calculate $\hat r^{\prime}$ we see that equation (\ref{newplanos}) is
\begin{equation}
\hat \nabla \chi \frac{\nabla F}{3F}=\frac{d \rho}{d \chi}(\hat \nabla  \chi )^2 ,
\label{Fro}
\end{equation}
with ${\displaystyle \rho\equiv \ln Z +\frac2Z}$. Since we can write (\ref{Fro}) as follows
\begin{equation}
\hat \nabla \chi \nabla f=\hat \nabla \rho \hat \nabla \chi \ , \ f\equiv \frac 13 \ln F \ ,
\label{fro}
\end{equation}
then, a general solution to this equation is $\rho=f+\mu$ for any function  $\mu(x,y)$ which verifies\footnote{As a matter of  curiosity the solutions of  equation (\ref{fro}) with the form $\rho=f+\mu$  can be understood as solutions of a Hamiltonian problem  with potential $V=-\frac 18 (\nabla f)^2$. Indeed, we can write  equation  (\ref{Fro}) not only as (\ref{fro}) but also as $
	\hat \nabla \rho \nabla f=(\hat \nabla \rho)^2
	$,
	since we can multiply both sides of  equation (\ref{Fro}) by the factor $\frac{d \rho}{d \chi}$ and consider $\frac{d \rho}{d \chi} \hat \nabla \chi=\hat \nabla \rho$. Next, by making a change of function $\hat \rho \equiv \rho -\frac f2$, we have $
	(\hat \nabla \hat \rho )^2=\frac 14 (\nabla f)^2
	$, that  has the trivial solution $\hat \rho=\frac f2$ , that is to say $\rho=f$, but again $\rho=f-g$ is another solution of the equation if $g$ is also a solution, and hence $\hat \rho=g-f/2=-\mu-f/2$ can be seen as another solution of the Hamiltonian problem.} $\nabla \mu \hat \nabla \chi=0$. Equivalently, we can say that the functions $\mu(x,y)$ and $Z$ are given by solving
\begin{equation}
F=Z^3 e^{\frac 6Z-3\mu} ,
\label{FFormat}
\end{equation}
$\mu(x,y)$ being a surface orthogonal to $Z$ : $\nabla \mu\hat \nabla Z=0$ (let us note that $\nabla \mu \hat \nabla \chi=0 \Leftrightarrow \nabla \mu \hat \nabla Z=0$).
Indeed, by solving $\mu$ from (\ref{FFormat}), ${\displaystyle \mu=\frac2Z-\frac 13\ln\left(\frac{F}{Z^3}\right)}$ and imposing  $\nabla \mu \hat \nabla Z=0$, we obtain the equation (\ref{Fro}), since $\hat \nabla Z=\frac{dZ}{d \chi}\hat \nabla \chi$, as follows:
\begin{equation}
\hat \nabla Z \frac{\nabla F}{3F}=\frac{d \rho}{d Z}(\hat \nabla  Z)^2.
\label{FZ}
\end{equation}

Before we  solve this equation, let us summarize the results obtained  by giving an answer to the {\it statement 2}. We figure out the following solution to that problem:

The MSA coordinates $\{\hat r, \hat y\}$ of the $MQ$-solution are (see equations (\ref{rdeZ}) and (\ref{ydeC}))
\begin{equation}
\hat r=\frac{a Z}{3 F}, \qquad    \hat y=\sqrt{\frac{C(x,y)+Q}{3 Q}}
\label{resume}
\end{equation}
$C(x,y)$ being  a solution of the orthogonality equation $\nabla \mu \hat \nabla Z=0$ with the constraint (\ref{FFormat}). In addition, this function, which is given by equations (\ref{CPi})-(\ref{Pi}), must lead to the verification of the  following asymptotic condition $\hat y(R\rightarrow \infty)=\omega$, $\hat r(R\rightarrow \infty)=R+M$.  Equivalently, from (\ref{Fro}) or (\ref{FZ}), the function  $C(x,y)$ is a solution, with the boundary behaviour and asymptotic conditions mentioned, of the following equation
\begin{equation}
\hat \nabla C \ \nabla F=\Omega (\hat \nabla  C)^2 \ , \ \Omega\equiv  \left\{
\begin{array}{cl}
 \frac{27 F^3(H^2-1)(H^2+1-H)}{2a^3 (H^2+1+H)^2\sqrt{\chi^2-1}}&,  \mid \chi \mid \geq 1 \nonumber\\
 & \nonumber\\
\frac{-27 F^3(1-\cos(2\varphi))\sin(\varphi)}{a^3(1+\cos(2\varphi))^2\sqrt{1-\chi^2}} & , \mid \chi \mid \leq 1,
\end{array}
\right.
\label{Ceq}
\end{equation}

In fact, (\ref{Ceq}) is the equation that we are forced to solve in our aim of looking for $C$, since the constraint (\ref{FFormat}) is a transcendent equation for $C$ and cannot be analytically solved. We make here the following  for solving  (\ref{Ceq}): let us  consider  a series expansion for $C$ in terms of the dimensionless quadrupolar parameter $q$,  as it was written in the expressions (\ref{CPi})-(\ref{Pi}). Whatever the value of $\chi$, either positive or negative one, the equation (\ref{Ceq}) with $C$ given by (\ref{CPi})-(\ref{Pi}) leads to a sequence of equations for each function $c_i$ at successive orders of $q$. These differential equations exclusively involves  derivatives with respect to the variable $x$ of the function $c_i$, because of the spherical character of $F$ at its order zero of the expansion in powers of $q$ ($F\sim \frac{2}{x+1}+O(q)$). And they must be solved choosing for  the corresponding constant of integration the suitable function of the angular variable $y$ required to fulfil the asymptotic conditions for  $\hat y$. The ordinary differential equations for the first two orders are the following\footnote{Let us note that the election made for the order $q$ of the function $C$ in (\ref{CPi})-(\ref{Pi}) justified by reasons of  spherical limit and boundary condition fulfilments leads to an identity at that order  $q$ in the differential equation.}
\begin{eqnarray}
&2& \frac{x-1}{x+1} c_2^{\prime} = \frac 32 a^3 (1-y^2)y\left( \frac 12 \dot{ f_1}-\frac{3y}{(x+1)^3}\right)  , \nonumber \\
&2& \frac{x-1}{x+1} c_3^{\prime} =(x^2-1)c_2^{\prime}f_1^{\prime}+ \nonumber \\
&+& a^3 (1-y^2)\left( \frac 34 y\dot{f_2}+\dot{c_2}\dot{f_1}-\frac{12 y}{(x+1)^3}\dot{c_2}-\frac{27 y^2f_1}{4 (x+1)^2}-\frac{27 y^2(1-3y^2)}{4 (x+1)^5} \right) , \nonumber\\
\label{esta}
\end{eqnarray}
where the following notation is used ${\displaystyle F=\frac{2}{x+1}+\sum_{i\geq 1}f_i(x,y)q^i}$, and the symbols $( ^{\prime} )$ and $( \dot{\ } )$ denote derivatives with respect to variables $x$ and $y$ respectively. The general solution for the first equation of (\ref{esta}) is
\begin{eqnarray}
c_2(x,y)&=&K(y)+\frac{3 a^3 y^2}{64}\left[ -3(1-y^2)(5x^3-5x+4)\ln\left(\frac{x-1}{x+1}\right)\right.+\nonumber \\
&+&\left.2\left(\frac{x-1}{x+1}\right)\left( -30x(1-y^2)+15x^2y^2\right)+\right. \nonumber\\
&+&\left.2\left(\frac{x-1}{x+1}\right)\left(\frac{9y^2(1-y^2)+24x^2y^2-15x^4+x^2+20x+10}{x^2-y^2}\right)\right] ,\nonumber\\
\label{lac2}
\end{eqnarray}
where $K(y)$ is the {\it constant} of integration. Now, the boundary behaviour of $c_2$ and the asymptotic condition for $\hat y$ are easily implemented: on the one hand $c_2(x,y=\pm1)=0$ in (\ref{lac2}) leads to $K(y)=-a^3 y^2 \frac{15}{16}+A(y)$, $A(y)$ being an arbitrary function which vanishes at the axis. On the other hand   we expand the angular variable $\hat y$ from (\ref{ydeces}) in terms of the quadrupolar parameter $q$ as follows ($\hat y\sim \hat y_0+\hat y_1 q+\hat y_2 q^2+\hat y_3 q^3+\cdots$)
 \begin{equation}
 \hat y\sim y+\frac{c_2}{6m^3y}q+\left[\frac{c_3}{6m^3y}-\frac{c_2^2}{72m^6y^3}\right]q^2+\left[\frac{c_4}{6m^3y}-\frac{c_2c_3}{72m^6y^3}+\frac{c_2^3}{432m^9y^5}\right]q^3+ \cdots
  \end{equation}
 and since the angular variable $y$ already verifies\footnote{Let us remember that $x=\frac{1}{2\lambda}(r_++r_-)$ and $y=\frac{1}{2\lambda}(r_+-r_-)$ with $r_{\pm}\equiv \sqrt{1+2\omega \lambda+\lambda^2}$ and $\lambda\equiv M/R$. Hence the asymptotic expansion $R\rightarrow \infty$ ($\lambda \rightarrow 0$)  leads to $x\sim 1/\lambda$  as well as $y\sim \omega$.} the asymptotic condition imposed ($\hat y(R\rightarrow \infty)=\omega$), we only need to require each $c_i$ to be  of order equal or higher than $O(1/R)$.
 This condition univocally  determines the function $A(y)$ as follows
\begin{equation}
K(y)=-a^3y^2\left(\frac{15}{16}+\frac{33}{32}(1-y^2)\right).
\end{equation}

\section{Relevant features}

\vspace{2mm}
$\bullet$ The  metric component $g_{00}^{MQ}$  looks like\footnote{This fact is actually the condition (i). From  the ansatz (\ref{ansatz}), whatever function $C$ verifies ${\displaystyle F=\frac{2M}{\hat r}+ \frac{C}{\hat r^3}}$ since (\ref{zetader}) is a solution of (\ref{ansatz}). In addition, the function $C(x,y)$ has been calculated to be (\ref{CP2Q}). } ${\displaystyle -1+\frac{2M}{\hat r}+\frac{2Q}{\hat r^3} P_2(\hat y)}$ in MSA coordinates. The case $Q=0$ leads to the spherical form ${\displaystyle -1+\frac{2M}{\hat r}}$ which is the Schwarzschild metric component $g_{00}^S$ because the MSA coordinates obtained (\ref{resume}) for the $MQ$-solution  has the appropriate limit $Q=0$, that is to say, the  Erez-Rosen coordinates:
\begin{equation}
\hat r({Q=0})=M(x+1) \qquad , \qquad \hat y({Q=0})=y
\end{equation}
since $\chi({Q=0})=H({Q=0})=1$, $Z({Q=0})=3$ and ${\displaystyle F({Q=0})=\frac{2}{x+1}}$.

\vspace{2mm}
$\bullet$ In \cite{msa} a first attempt of calculating  $MSA$ coordinates was done. As we have explained, an asymptotic expansion of $\{\hat r, \hat y\}$ was performed. Now, our expressions (\ref{resume}) recover exactly  that asymptotic behaviour ($l\equiv M/R$) as follows (see Appendix B in \cite{msa}),  ($\hat y\sim \hat y_0+\hat y_1 q+\hat y_2 q^2+\hat y_3 q^3+\cdots$):
\begin{eqnarray}
\hat y_0&=&y\sim \omega\left[ 1+\frac 12(\omega^2-1)l^2+\left(\frac 78\omega^4-\frac 54\omega^2+\frac 38\right)l^4 \right]+\cdots\nonumber \\
\hat y_1&=&\omega (\omega^2-1)\left[ l^3+\left(\frac{35}{28} \omega^2-\frac{39}{28} \right) l^4+\left(3 \omega^2-\frac {4}{5}\right)l^5\right]+\cdots \nonumber
\end{eqnarray}
\begin{eqnarray}
\hat r_0&=&M(x+1) \sim \frac Ml+M-\frac M2(w^2-1)l-\frac M8(\omega^2-1)\left(5\omega^2-1\right)l^3+\cdots \nonumber \\
\hat r_1&=&\frac 12 M(1-3\omega^2)l^2+M\left(-\frac54 \omega^4+\frac{39}{14} \omega^2-\frac{19}{28}\right) l^3+M\omega^2\left(2-3\omega^2\right)l^4+\cdots \nonumber \\
\hat r_2&=&M\left(12\omega^4-9\omega^2+1\right)l^4-\frac{1}{616}M(1155\omega^6+13089\omega^4-10347\omega^2+1247)l^5+\cdots \nonumber \\
\end{eqnarray}

\vspace{2mm}
$\bullet$ While it is true that the function $C$  is still written in terms of a series (\ref{CPi})-(\ref{Pi}), it is likewise so that the functions $c_i(x,y)$ of that series are characterized by  an ordinary  differential equation, and hence we are able to determine its general solution by means of a boundary condition and asymptotic behaviour.

Of course that equation (\ref{resume}) does not provide us with an exact expression for the  MSA coordinates, but they can be considered as approximative as we like since the quadrupolar parameter $q$ is small ($\arrowvert q \arrowvert<<1$), and we could neglect terms of high order in $q$ depending on the accuracy desired for the use we were making of these coordinates.

It should be emphasized that if the sum of the series (\ref{ydeces}) for the function $C$ were known then it should result in the achievement of the sum of the series for the metric function $\Psi$ of the $MQ$-solution\footnote{In \cite{mq} the  Weyl family of solutions ${\displaystyle \sum_{n=0}^{\infty} a_n\frac{P_n(\omega)}{R^{n+1}}}$ was converted into  the functions series (\ref{sumaton}) in the parameter $q$ for the $MQ$-solution (see also \cite{tesis})}
\begin{equation}
\Psi_{MQ}=\Psi_0+\Psi_1 q+\Psi_2 q^2+\cdots = \frac 12 \ln\left(1-\frac{2M}{\hat r}- \frac{2Q}{\hat r^3} P_2(\hat y)\right) .
\label{sumaton}
\end{equation}

\vspace{2mm}
$\bullet$ In \cite{event} the event horizon of the $MQ$-solution was described, assuming that  the $MSA$ coordinates of that solution were known. The only  \cite{israel} static and asymptotically flat vacuum space-time possessing a regular horizon is the Schwarzschild solution. In fact, all the other Weyl exterior solutions exhibit singularities in the physical components of the Riemann tensor at the horizon. Nevertheless, we can study for any static metric different from the Schwarzschild solution, the surface of infinite redshift defined by the condition $g_{00} \equiv \xi^{\alpha} \xi_{\alpha} = 0$ ($\xi^{\alpha}$ being the time-like Killing vector). In those cases, the deviation of the solution from the spherical symmetry leads to a surface different to $r=2M$ surface of Schwarzschild (event horizon). Although  strictly speaking the term 'horizon' refers to the spherically symmetric case, we shall use it when considering the {\it $r = 2M$ surface}, in the case of small deviations from sphericity.

 The condition $g_{00}=0$ leads to the surface $\hat r=\hat r(\hat y)$ of infinite redshift where the radial coordinate $\hat r$ is  a function of $\hat y$, and the deformation with respect to the Schwarzschild surface produced by the quadrupole moment can be shown, preserving a regular, closed, continuous and differentiable surface.  Now, since we have obtained  the radial coordinate $\hat r$ as a function of $F\equiv 1+g_{00}$, $\hat r=\hat r(F,2QP_2(\hat y))$, we are able to conclude  that the surface of the event horizon of the $MQ$-solution corresponds to $\hat r(F=1)$ in  (\ref{rdeZ}). (Note that $F=1$ is not $F_1$ in the notation used since $F_1$ stands for the value of $F$ at the symmetry axis). In fact, we recover the expression for the event horizon obtained in \cite{event}. But in addition we are able to correct a mistaken conclusion derived in that paper: let us remind that the continuity of the piecewise surface (\ref{rdeZ}) is fulfilled at $\chi=1$, and  upper and lower bounds can be established for $\chi$,  $1-\frac{27}{16} < \chi <1+\frac{27}{8}$. Nevertheless these limiting values are not related with bounds in the parameter $q$ as it was concluded in \cite{event} but the range  of values  for $\chi$ exclusively derives from the extreme values of the Legendre polynomial $P_2 \in[-1/2,1]$ and the assumption that the quadrupolar parameter $q$ is small ($\arrowvert q \arrowvert<1$). Hence we have to clarify the unphysical and unexplained conclusion  showed in \cite{event}, quite the opposite  there is no restriction to the value of the quadrupolar parameter except for $\arrowvert q \arrowvert<1$.

\vspace{2mm}
$\bullet$ Could we make use of this procedure to find any solution of the Weyl family with a finite number of RMM by means of looking for their corresponding MSA coordinates? I mean, let us suppose that we handle with  a general solution of the Weyl family, assuming deliberately  unknown Weyl coefficients $a_n$, for instance for  the $MQ$-solution, and we develop the procedure already set out aiming to find the MSA coordinates of this solution. The question raised is whether the procedure leads to the determination of the  Weyl moments $a_n$ of the solution. First, we have to note that, without the explicit knowledge of the Weyl coefficients, the metric component $g_{00}$ as well as the function $C$ are   series in the inverse of the radial coordinate $R$. The application of the procedure leads to relationships between the functions $\hat c_i$ (different from the $c_i$ used before) and the coefficients $a_n$, but at the end the result is similar than the one obtained  in \cite{msa}  by using an asymptotic expansion of the MSA coordinates.  Secondly, a freedom in the election of the $a_n$ coefficients exits which cannot be fixed by means of some asymptotic condition.

In conclusion we find that,  for a given set of $a_n$ Weyl coefficients an unique system of  MSA coordinates exists which leads  to vanishing Thorne's rests, but on the contrary the MSA coordinates by itselves cannot univocally determine  the Weyl coefficients of the solution.

Therefore this point addresses relevance to the knowledge of the $MQ$-metric in terms of the quadrupolar parameter \cite{mq}, because it allows us to determine, as we have shown, the MSA coordinates by means of solving  a Dirichlet problem.

\vspace{2mm}

$\bullet$ The procedure explained can be extended to the general case of any Weyl solution with a finite number of RMM. Let us consider, for instance,  the axially symmetric and static solution possessing only mass, quadrupole moment $Q$ and $2^4$-pole moment $M_4$ (massive RMM). If we would know the metric component $g_{00}$  ($F\equiv g_{00}+1$) of that solution we would look for MSA coordinates $\{\hat r, \hat y\}$ such that
\begin{equation}
F=\frac{2M}{\hat r}+\frac{2Q}{\hat r^3} P_2(\hat y)+\frac{2M_4}{\hat r^5} P_4(\hat y)
\label{Fm4}
\end{equation}
From this equation which is relating  the angular variable $\hat y$ and the radial $\hat r$, the orthogonality condition of the coordinates leads to the  following equation, equivalent to (\ref{3rF})
\begin{equation}
\left(-5F\hat r^3+8M\hat r^2+4QP_2(\hat y)\right)(\nabla \hat r)^2=\hat r^4 \nabla F \cdot \nabla \hat r .
\label{5Fm4}
\end{equation}

Now, we assume that $\hat r=\hat r(\zeta,\{x,y\})$ and analogously to the previous case we consider the ansatz $\left(-5F\hat r^3+8M\hat r^2+4QP_2(\hat y)\right)\hat r^{\prime}=\hat r^4$ to solve equation  (\ref{5Fm4}) as follows
\begin{equation}
\zeta=F=\frac{2M}{\hat r}+\frac{2Q}{\hat r^3} P_2(\hat y)+\frac{C}{\hat r^5},
\label{zetaF}
\end{equation}
$C=C(x,y)$ being  a function of the coordinates which must be  equal to  $2M_4$ on the symmetry axis. In fact, from (\ref{zetaF}) and (\ref{Fm4}) we see that $C=2M_4P_4(\hat y)$ and hence
\begin{equation}
\hat y^2=\frac 37\left[1\pm\frac 23\sqrt{\frac{6M_4+7C}{5M_4}}\right] .
\label{ym4}
\end{equation}
The requirement of preserving the symmetry axis i.e. $\hat y(y=\pm1)=y=\pm1$ forces to take the positive values  in  (\ref{ym4}) and we consider for $C$ the following expression with $c_i(y=\pm1)=0$ and $m_4\equiv M_4/M^5$
\begin{equation}
C=2M_4+\sum_{i=1}^{\infty}c_i m_4^i .
\label{cm4}
\end{equation}

Therefore, the radial coordinate $\hat r$  for this case is given by a real root of the equation (\ref{zetaF}), and the angular variable $\hat y$ is
\begin{equation}
\hat y^2=\frac 37\left[1+\frac 23\sqrt{4+\frac{7}{5M^5}\sum_{i=1}^{\infty}c_i m_4^{i-1}}\right] .
\label{ym4def}
\end{equation}

The equation for the function $C=C(x,y)$ is again (\ref{newplanos}) which has to be solved within the asymptotic condition $\hat y(R\rightarrow \infty)=\omega$. This equation (\ref{newplanos}) requires the explicit expression for the radial coordinate $\hat r=\hat r(\zeta,C)$ and from the new ansatz and (\ref{zetaF}) we have that
\begin{equation}
-\frac{1}{\hat r^{\prime}}=\frac{5F}{\hat r}-\frac{4QP_2(\hat y)}{\hat r^4}-\frac{8M}{\hat r^2} .
\end{equation}

The problem of looking for the root of algebraic equation (\ref{zetaF}) can be solved in terms of A-hypergeometric series in the sense of
Gelfand, Kapranov and Zelevinsky \cite{zeles} as showed in  \cite{sueco} (and references therein)

\vspace{2mm}
$\bullet$ The relevance of slight deviations from spherical symmetry is a topic widely treated in literature (see for example \cite{mq},\cite{tesis},\cite{RF2}-\cite{RF7}, and references therein.) And there exist possibilities to make astrophysical applications of these solutions (see for example \cite{RF6}, \cite{RF7}) in terms of the RMM of the solutions considered. The influence of the quadrupole moment on the motion of test particles within the context of the Erez-Rosen metric \cite{R13} has been investigated by many authors (see \cite{R14}-\cite{R17} and references therein). Some other works have been devoted to obtain relations between the RMM and observables  trying to establish mechanisms to {\it measure} them.
In \cite{R1} we calculated, by using the technique of Rindler and Perlick,  the total precession per revolution of a gyroscope circumventing the source of Weyl metrics. We established thereby a link between the multipole moments of the source and an {\it observable} quantity.

In \cite{R2} we analyzed the behavior of the geodesic motion of test particles in the space-time of an specific class of axially symmetric static vacuum solutions to the Einstein equations, the linearized multipole solution \cite{R3}. Similar study was done in  \cite{RF7}. The existence of an innermost stable circular orbit very close to the (singular) horizon of the source is established. The existence of such a stable orbit, closer than that of the Schwarzschild metric, as well as the appearance of a splitting in the admissible region of circular orbits, is shown to be due to the multipole structure of the solution, thereby providing additional potential observational evidence for distinguishing Schwarzschild
black holes from naked singularities.

In \cite{R4} a static and axisymmetric solution of the Einstein vacuum equations with a finite number of RMM is written in multipole symmetry adapted (MSA) \cite{mq} coordinates up to certain order of approximation. From the equation of equatorial geodesics, we obtain the Binet equation for the orbits and it allows us to determine the gravitational potential that leads to the equivalent classical orbital equations of the perturbed Kepler problem. The relativistic corrections to Keplerian motion are provided by the different contributions of the RMM of the source starting from the monopole (Schwarzschild correction). In particular, the perihelion
precession of the orbit is calculated in terms of the quadrupole and $2^4$-pole moments. Since the MSA coordinates generalize the Schwarzschild coordinates, the result obtained allows measurement of the relevance of the quadrupole moment in the first order correction to the perihelion frequency-shift.

Therefore, the problem that we have developed in this work, aiming to determine the MSA coordinates, leads to an scenario where the quadrupole parameter plays a relevant role that allows to show once again the enormous differences arising when leaving the spherical symmetry case.

\section{Conclusions}

We have constructed a system of coordinates (\ref{resume}) with which the Thorne's rests of the metric function $g_{00}$ vanish. Hence, the relativistic exterior gravitational field of an isolated compact source with a finite number of RMM can be described by means of a function ${\displaystyle u\equiv \frac 12 (1+g_{00})=\frac 12 (1-e^{2\Psi})}$ that resembles the form of the Newtonian gravitational potential. The particular case of the $MQ$-solution of the  static and axisymmetric  Einstein vacuum equations is developed, as well as a wide description of the general case is discussed. The coordinates obtained for the $MQ$-solution are the so-called MSA coordinates and we recover the asymptotic expansion calculated for that system of coordinates in \cite{msa}.

In addition, the system of coordinates obtained is characterized as the unique solution of a differential equation with a boundary and asymptotic conditions. We consider this system of coordinates as a quadrupolar generalization of the Erez-Rosen system and in fact we recover it from (\ref{resume}) when the quadrupole parameter vanish, for the limit case $q=0$.

These coordinates allow us to calculate the event horizon of the $MQ$-solution, in a simply way, by taking the function $F$ equal to $1$ in the expression of the radial coordinate. And finally, it is remarkable that the MSA system of coordinates is relevant because it is linked to a group of symmetries of the Laplace and Ernst equations, and the solutions with a finite number of multipoles can be defined as their group invariant solutions \cite{msa}.

\section{Appendix 1}

\noindent {\bf Theorem}

Let $\Lambda$ be a function such that  $\Lambda=\Lambda(\beta)$, $\beta$ being an arbitrary function. The equation
\begin{equation}
H(\Lambda) \triangle \Lambda - \frac{dH}{d\Lambda}(\nabla \Lambda)^2=0
\label{ap1}
\end{equation}
is equivalent to $\triangle \beta=0$ {\it iff} $H=\delta \frac{d \Lambda}{d \beta}$, for any constant $\delta$.

\vspace*{5mm}

\noindent {\it Proof}:

Since ${\displaystyle \triangle \Lambda=\frac{d^2\Lambda}{d \beta^2}(\nabla \beta)^2+\frac{d\Lambda}{d\beta}\nabla \beta}$,  then the equation (\ref{ap1}) is
\begin{equation}
H \frac{d\lambda}{d\beta}\triangle \beta+(\nabla\beta)^2\left[H\frac{d^2\Lambda}{d\beta^2}
-\frac{dH}{d\Lambda}\left(\frac{d\Lambda}{d\beta}\right)^2\right]=0 .
\label{ap2}
\end{equation}
Hence, (\ref{ap2}) is equivalent to $\triangle \beta=0$ iff
\begin{equation}
H\frac{d^2\Lambda}{d\beta^2}
=\frac{dH}{d\Lambda}\left(\frac{d\Lambda}{d\beta}\right)^2 ,
\end{equation}
whose solution is ${\displaystyle H=\delta\frac{d\Lambda}{d\beta}}$ for any constant $\delta$. $\hfill{\square}$

This theorem allows us to solve the kind of equations (\ref{ap1}) by means of an auxiliary function $\beta$ which is harmonic. Let us see some examples:

{\it Example 1}.-  The Ernst equation, $E \triangle E-(\nabla E)^2=0$, admits the following solution: $E=e^{2\eta}$  whatever harmonic function $\eta$, $\triangle \eta=0$.

{\it Example 2}.-  The transformed Ernst equation, $(\xi^2-1) \triangle 2\xi-(\nabla \xi)^2=0$, admits the following solution: ${\displaystyle \xi=\frac{1-e^{2\eta}}{1+e^{2\eta}}}$  whatever harmonic function $\eta$, $\triangle \eta=0$.

\section{Appendix 2}

The general solution of equation (\ref{planos}) is $\zeta=F-G$, $G$ being another solution of the same equation. Then, the ansatz (\ref{ansatz}) is now
\begin{equation}
(-3(\zeta+G) \hat r+4 M) \hat r^{\prime}=\hat r^2 ,
\label{ansatzconG}
\end{equation}
and it has the following solution
\begin{equation}
\zeta=\frac{2M}{\hat r}-\frac 32 \frac{G}{\hat r}+\frac{C}{\hat r^3}=F-G .
\label{FmenosG}
\end{equation}
Hence, the radial coordinate $\hat r$ gives
\begin{equation}
\hat r=\frac{\hat aZ(\hat \chi)}{3\zeta},
\end{equation}
with $Z$ defined in (\ref{rdeZ}) but now the parameter $a$ of (\ref{rdeZ}) becomes a function $\hat a\equiv 2M-\frac 32 G$, and $\chi$ becomes  ${\displaystyle \hat \chi=\frac{27}{2}\frac{C(F-G)^2}{\hat a^3}+1}$.

We handle now with two functions $C$ and $G$ which are connected between them by  equation (\ref{FmenosG}) ${\displaystyle F=\frac{\hat a}{\hat r}+\frac{C}{\hat r^3}+G}$ and the condition (i) of  the {\it statement 1}.   It is not necessary to keep taking both functions and, without loss of generality, we can consider $C$ equal to a constant $C=2Q$ and $G$ be demanded to fulfill the boundary condition on the axis $G(y=\pm1)=0$.
Hence, the resulting MSA coordinates are given by
\begin{equation}
\hat r=\frac{\hat aZ}{3\zeta} , \qquad \hat y=\sqrt{1-\frac{G \hat r^2}{6Q}(3-2 \hat r)} .
\label{msaG}
\end{equation}
Now,  the radial coordinate $\hat r=\hat r(\zeta,\{x,y\})$ depends on the coordinates $\{x,y\}$ through the function $G$, and the equation that it  must satisfy is analogous to (\ref{newplanos}):
\begin{equation}
\hat \nabla \hat r \cdot \nabla (F-2G) =-\frac{1}{\hat r^{\prime}}(\hat \nabla \hat r)^2-\hat r^{\prime} \left[ (\nabla G)^2-\nabla G \cdot \nabla F\right] .
\end{equation}

The resulting equation for $G$ is the following
\begin{equation}
\left[ 1-\frac{\hat r^{\prime}}{\epsilon}\right] \nabla G \cdot \nabla F+\left[ -2+\frac{\hat r^{\prime}}{\epsilon}+\frac{1}{\hat r^{\prime}}\right](\nabla G)^2=0 ,
\label{G}
\end{equation}
with
\begin{equation}
\frac{1}{\hat r^{\prime}}=-\frac{9\zeta^2 H(H^2+1-H)}{a(H^2+1+H)^2} , \quad \epsilon\equiv -\frac 32 \frac{H^2-1}{3\zeta}\frac{H^3-1}{\sqrt{\chi^2-1}} .
\label{1sobrerprima}
\end{equation}

The complexity of the equation for $G$ and the fact that the resulting angular coordinate $\hat y$ depends on $\hat r$ again (see equation (\ref{msaG})) make reasonable to consider the case $G=0$ as well as $C=C(x,y)$ as it has been done in the  sections of the paper.

\section{Acknowledgments}
This  work  was partially supported by the Spanish  Ministry of Science and Innovation
under Research Project No. FIS2015-65140-P (MINECO/FEDER), and the Consejer\'\i a de Educaci\'on of the Junta de Castilla y
Le\'on under the Research Project SA083P17, Grupo de Excelencia GR234 and UIC.

Let this be a humble tribute to my great Teacher Chus Martin from a grateful scientific son.


\begin{thebibliography}{88}
	
	
		\bibitem{msa} Hern\'andez-Pastora, J.L. (2010)
	{\it Class. Quantum Grav.} {\bf 27}, 045006 (20pp).
	
	\bibitem{mq} Hern\'andez-Pastora, J.L.,  Mart\'\i n, J. (1994)
	{\it Gen. Rel. and Grav..} {\bf 26}, 877.
	
	
		\bibitem{event} Hern\'andez-Pastora, J.L. and L. Herrera (2011)
	{\it Class. Quantum Grav.} {\bf 28}, 225026 (14pp)
	
	
		\bibitem{weyl}  Weyl, H. (1917) {\it Ann. Phys.} (Leipzig) {\bf 54}, 117
		
		
		
		\bibitem{tesis} Hern\'andez-Pastora, J.L., (1996)
		{Ph.D. Relativistic gravitational fields close to Schwarzschild solution}. Universidad de Salamanca.
		
	
	\bibitem{ernst} Ernst, F.J., (1968) {\it Phys. Rev.}, {\bf 167}, 1175,
	
	Ernst, F.J.. (1968) {\it Phys. Rev.}, {\bf 168}, 1415.
	
	
		\bibitem{Nsym} Hern\'andez-Pastora. (2008)
	{\it Class. Quantum Grav.} {\bf 25}, 165021 (21pp).
	
	
		\bibitem{varios} Hern\'andez-Pastora, J.L.,  Mart\'\i n, J. (1993)
	{\it Class. Quantum Grav.} {\bf 10}, 2581.
	
	Hern\'andez-Pastora. (2006)
	{\it Gen. Rel. and Grav.} {\bf 38}, 871.
	
	Hern\'andez-Pastora, J.L.,  Mart\'\i n, J and E. Ruiz (1998)
	{\it Gen. Rel. and Grav.} {\bf 30}, 999.
	
	
	
\bibitem{sueco}	B\"ackdahl, T., Herberthson,  M., (2005) {\it Class. Quantum Grav.} {\bf 22}, 1607-1621.

	
	\bibitem{geroch} Geroch, R. (1970) {\it J.  Math. Phys.}, {\bf 11}, 2580.
	
		\bibitem{thorne} Thorne, K.S., (1980) {\it Rev. Mod. Phys.}, {\bf 52}, 299.
	

	
	\bibitem{prolates} H. Quevedo, (1990) {\it Fortschr. Phys.} {\bf 38}, 10, 733-840.
	
	
	
	
\bibitem{zeles} Gel’fand I M, Zelvinsky A V and Kapranov M M, (1989) {\it	 Anal. Appl.}, {\bf 23}, 94

	
	
	
		
\bibitem{israel} Israel W (1967) {\it Phys. Rev.} {\bf 164},  1776
	

	\bibitem{R1} Herrera, L. and Hernandez-Pastora, J.L. (2000) {\it J. Math. Phys.} {\bf 41}, 7544
	
\bibitem{R2}  Hernandez-Pastora, J.L., Herrera, L.  and Ospino, (2013) {\it J. Phys. Rev. D} {\bf 88}, 064041

\bibitem{R3} Class. Quantum Grav. 30 (2013) 175003 (21pp)

\bibitem{R4}  Hernandez-Pastora, J.L. and Ospino,(2010) {\it J. Phys. Rev. D}{\bf  82}, 104001

\bibitem{R6} Winicour J, Janis A I and Newman E T(1968) {\it Phys. Rev.} {\bf 176},  1507

\bibitem{R7} Janis A I, Newman E T and Winicour J (1968) {\it Phys. Rev. Lett.} {\bf 20},  878

\bibitem{R8} Cooperstock F I and Junevicus G J (1973) {\it Nuovo Cimento B} {\bf 16},  387

\bibitem{R9} Bel L (1971) {\it Gen. Rel. Grav.} {\bf  1},  337

\bibitem{R10} Herrera L (2008) {\it Int. J. Mod. Phys. D} {\bf 17},  557

\bibitem{R11} Virbhadra K S and Ellis G F R (2002) {\it Phys. Rev. D} {\bf 65},  103004

\bibitem{R12} Virbhadra K S and Keeton C (2008)  {\it Phys. Rev. D} {\bf 77}, 124014

\bibitem{R13} Erez G and Rosen N (1959) {\it Bull. Res. Council Israel F} {\bf  8} 47

\bibitem{R14} Zeldovich Ya and Novikov I D (1971) {\it Relativistic Astrophysics} (Chicago, IL: University of Chicago Press)

\bibitem{R15} Armenti A and Havas P (1971) {\it Relativity and Gravitation ed C Kupper and A Peres} (London: Gordon and Breach)

\bibitem{R16} Quevedo H (1990) {\it Fortschr. Phys.} {\bf 38}, 733

\bibitem{R17} Mashhoon B and Quevedo H (1995) {\it Nuovo Cimento B} {\bf  110},  291




\bibitem{RF2}  Voorhees B.H.,(1970) {\it J. Phys. Rev. D} {\bf  2}, 2119

\bibitem{RF3}  Quevedo H.,(1986) {\it J. Phys. Rev. D} {\bf  33}, 324

\bibitem{RF4}  Quevedo H.,(1987) {\it Gen. Rev. Grav.} {\bf  19},1013

\bibitem{RF5} Quevedo H.,(1989) {\it J. Phys. Rev. D} {\bf  39}, 2904

\bibitem{RF6} Gibbons G.W. and Volkov M.S. (2017) {\it J. Cosmol. Astropart. Phys.} {\bf 05},  039

\bibitem{RF7} Turimov B., Ahmedov B., Kolos M. and Stuchlik Z. (2018) {\it Phys. Rev. D} {\bf 98}, 084039

	
\end{thebibliography}
\end{document}